\documentclass[12pt]{article}
\newlength{\vshift}
\input epsf.tex
\newlength{\hshift}

\usepackage{amssymb}
\usepackage{color}
\usepackage{amsmath}
\usepackage{amsthm}
\usepackage{amsopn}
\usepackage{axodraw}
\usepackage{graphicx}
\def\uno{\mbox{1 \kern-.59em {\rm l}}}

\def\th{\theta}
\def\nn{\nonumber}
\def\be{\begin{equation}}
\def\ee{\end{equation}}
\def\bea{\begin{eqnarray}}
\def\eea{\end{eqnarray}}

\begin{document}

 \vspace*{0cm}

 \begin{center}

  {\bf{\large Annihilation of singlet fermionic dark matter into two photons
  }}

\vskip 4em

 {{\bf M.M. Ettefaghi} \footnote{ mettefaghi@qom.ac.ir} and {\bf R. Moazzemi}\footnote{r.moazzemi@qom.ac.ir}
 }
 \vskip 1em
 Department of Physics, The University of Qom, Qom 371614-611,
Iran.

 \end{center}

 \vspace*{1.9cm}

\begin{abstract}
We consider an extension of the standard
model in which a singlet fermionic particle, to serve as cold dark
matter, and a singlet Higgs are added. We perform a reanalysis on the free parameters. In particular, demanding a correct relic abundance of dark matter, we derive and plot the coupling of the singlet fermion with the singlet Higgs, $g_s$, versus the dark matter mass. We analytically compute the
pair annihilation cross section of singlet fermionic dark matter
into two photons. The thermally averaged of this cross section is calculated for wide range of energies  and plotted versus dark matter mass using $g_s$ consistent with the relic abundance condition. We also compare our results with the Fermi-Lat observations.

\end{abstract}
PACS:
\newpage
\section{Introduction}
Although there exist numerous cosmological and astrophysical
indications which confirm the presence of the non-baryonic dark
matter, the nature of this substance remains unknown
\cite{rev1,rev2}. One of the most important motivations for the
extension of the standard model (SM) is to introduce an appropriate
candidate for dark matter. Among the various candidates proposed by
the theories beyond the SM, weakly interacting massive
particles (WIMP) are the most popular one. For instance
supersymmetry models with R parity \cite{rev1,rev2}, the extra
dimensional models with conserved Kaluza-Klein (KK) parity
\cite{kk}, the T-parity conserved little Higgs model \cite{little
Higgs}, and so on provide a WIMP dark matter candidate. Meanwhile,
these models introduce others degrees of freedom besides of the dark
matter which have not been confirmed experimentally until now. Of
course, in addition to the dark matter there exist others
motivations
 such as neutrino oscillation, baryon
asymmetry, hierarchy problem, and so on for considering such models.
However, maybe it seems natural to consider the most economical
model for the explanation of these anomalies. For instance, if we
want to explain dark matter anomaly, adding a singlet scalar to the
SM content to play the dark matter role is the most
minimal model. Several authors have studied the singlet scalar dark
matter \cite{scalar}. Alternatively, adding a singlet fermion together with a
singlet Higgs to the SM content leads to a renormalizable
theory for explanation of the dark matter \cite{fermion1}. Losing
the renormalizablity of the theory we can have more minimal model
with singlet fermionic dark matter (SFDM) \cite{fermion2,q}. In all above
scenarios, the weak interactions of WIMPs are the main key for
explaining the thermal production of them in the early universe (for
review see \cite{rev1,rev2}). Additionally, these weak interactions
can provide an opportunity for search dark matter candidates through their
production in high energy accelerators \cite{peskin}, their direct
detection \cite{dama}, and their indirect detection i.e.
astrophysical observations of the annihilation or decay products of
the dark matter in our galaxy or beyond.

In fact, the dark matter anomaly is a mutual problem between
particle physics and astrophysics; on one hand, we need to know
particle physical properties of dark matter to attribute a galaxy
distribution to it and on the other hand we need to know the
distribution of the dark matter to construe a direct or indirect
detected signals as  dark matter signals.

In the WIMP scenario, the weak interaction of dark matter would
produce observable SM particles, such as charged
anti-matter particles, photons  and neutrinos. Among these, neutrinos
and photons have advantage in comparison to others because they keep
their source information during the streaming. Moreover, the very
small cross sections of the neutrinos make their flux very difficult to
detect. Therefore, the gamma ray signature of the dark matter has
been investigated extensively (For review see \cite{gamma rev} and
references therein). Since the continuum gamma ray emission from dark
matter annihilation could be confused with astrophysical backgrounds,
e.g. emission from galactic cosmic rays or millisecond
pulsars, the study of the monochromatic gamma ray is important.
Monochromatic gamma ray signatures have been studied for some dark
matter candidates in the literature \cite{gammaray}. In this paper we
investigate monochromatic gamma ray in the context of  the SFDM model.

The most minimal model, which includes a singlet fermion to serve as
dark matter, is discussed in \cite{fermion2}. In this model the
leading interaction between dark matter and the SM
particle is given by the dimension-five term
$\frac{1}{\Lambda}H^\dag H\bar{\psi}\psi$. Here  $H$, $\psi$ and
$\Lambda$ are the SM Higgs doublet, the
dark matter fermion and a new physics scale, respectively. Moreover, it is
possible to explain the dark matter production in the early universe
by a renormalizable theory if the Higgs sector of the SM is extended
by a new singlet Higgs \cite{fermion1}. Also in the noncommutative
space-time we can have a singlet fermionic dark matter which is involved
in the noncommutative induced $U(1)$ gauge theory \cite{q}. There are also some others phenomenological studies
on this model \cite{damafermion}.


In this paper, we reobtain the thermal average of the annihilation of SFDM into the SM particles. Then, demanding the correct relic abundance
 we derive the coupling of SFDM with the singlet Higgs versus dark matter mass. For this calculation we use a set of Higgs masses so that
the recent CMS and ATLAS results are respected \footnote{http://cms.web.cern.ch/news/observation-new-particle-mass-125-gev}. Then we obtain the cross section of the
annihilation of SFDM pairs into two photons. The thermal average of this cross section is calculated throughout a favorite WIMP mass interval including resonance region. We depict the SFDM annihilation cross section into two photons versus dark matter mass, $m$, using $g_s$ consistent with relic abundance condition. We have organized the paper as
follows: In the next section, we review the renormalizable extension
of the SM by a singlet fermion which can be served as
a cold dark matter and derive the $g_s$ consistent with the relic density abundance. In Section \ref{sec3}, we calculate the two photons
annihilation of the singlet fermion pairs and its thermally averaged cross section. We summarize our result in the last section.
\section{Renormalizable Extension of the Standard Model by a Singlet Fermion}
It is possible to consider a singlet Dirac fermion beyond the SM
which can serve as cold dark matter. The lepton number and baryon
number of this new fermion are taken to be zero. Hence, there is not
any mixing between the new fermion and the SM fermions.
If a singlet Higgs is added to this model, the interactions between
the singlet fermion and SM particles can be explained by
a renormalizable manner. Explicitly, the singlet fermion and singlet
Higgs are coupled as follows:
 \be {\cal L}_{\rm{int}}=g_s\bar{\psi}\psi
S,\ee
 and the singlet Higgs can be coupled with the SM
Higgs through the following terms:
\be \label{hs}{\cal L}_{\rm{SH}}=-\lambda_1H^\dag HS-\lambda_2H^\dag
HS^2.\ee
Therefore, in
addition to the above terms, we need to add the following terms for singlet fermion
and singlet Higgs to the SM Lagrangian:
\be {\cal L}_\psi=\bar{\psi}(i\partial\!\!\!/-m_\psi)\psi,\ee
\be\label{selfS}{\cal L}_{\rm{S}}=\frac 1 2 (\partial_\mu S)(\partial^\mu
S)-\frac{m_0^2}{2}S^2-\frac{\lambda_3}{3!}S^3-\frac{\lambda_4}{4!}S^4.\ee
The scalar potential given in (\ref{hs}) and (\ref{selfS}) together
with the SM potential, $-\mu^2H^\dag H+\lambda_0(H^\dag
H)^2$, lead to the vacuum expectation values
\be \langle
H\rangle=\frac{1}{\sqrt{2}}\left(
                             \begin{array}{c}
                               0 \\
                               v_0 \\
                             \end{array}
                           \right) \qquad \mbox{and}\qquad \langle S\rangle=x_0
\ee
for the SM Higgs doublet which gives rise to the electroweak
symmetry breaking and for the singlet scalar
Higgs, respectively. These quantities are written with respect to the parameters
introduced in the Lagrangian by the following relations derived from
the extremum conditions, $\frac{\partial V}{\partial H}\left|_{{\langle
H\rangle=\frac{v_0}{\sqrt{2}}}}\right.=0$ and $\frac{\partial V}{\partial S}\left|_{\langle
S\rangle=x_0}\right.=0$:
 \bea
&&\mu^2=\lambda_0v_0^2+(\lambda_1+\lambda_2x_0)x_0,\nn\\
&&m_0^2=-\frac{\lambda_3}{2}x_0-\frac{\lambda_4}{6}x_0^2-\frac{\lambda_1v_0^2}{2x_0}-\lambda_2v_0^2.\eea
We define the scalar fields $h$ and $s$ as the deviation from the vacuum expectation values of $H$ and $S$,
respectively.
As it is obvious these fields are not mass eigenstates. The
various components of the mass matrix are written as follows:
\bea
&&\mu^2_h\equiv\frac{\partial^2V}{\partial
h^2}\Big{|}_{h=s=0}=2\lambda_0v_0^2,\nn\\
&&\mu^2_s\equiv\frac{\partial^2V}{\partial
s^2}\Big{|}_{h=s=0}=\frac{\lambda_3}{2}x_0+\frac{\lambda_4}{3}x_0^2-\frac{\lambda_1v_0^2}{2x_0},\nn\\
&&\mu^2_{hs}\equiv\frac{\partial^2V}{\partial h\partial
s}\Big{|}_{h=s=0}=(\lambda_1+2\lambda_2x_0)v_0.
\eea
The mass
eigenstates $h_1$ and $h_2$ are
\bea h_1=\sin(\th) s+\cos(\th) h,\nn \\
h_2=\cos(\th) s-\sin(\th) h, \eea
where the mixing angle $\th$ is defined by
\be \tan(\th)\equiv\frac{y}{1+\sqrt{1+y^2}}, \ee
with $y=\frac{2\mu^2_{hs}}{(\mu^2_h-\mu^2_s)}$. The Higgs boson
masses $m_1$ and $m_2$ are given by \be
m^2_{1,2}=\frac{\mu_h^2+\mu_s^2}{2}\pm\frac{\mu_h^2-\mu_s^2}{2}\sqrt{1+y^2}.
\ee This extension of the SM introduces some
new parameters. In the sense of experiment,
 there exist  allowed regions in the parameters space where it is possible to explain the
 relic abundance of the singlet fermions so that they serve as cold dark matter \cite{fermion1}.
Before the study of monochromatic gamma ray, we give a reanalysis of the relevant parameters. At the tree level of perturbation, the annihilation  of the singlet fermions into the standard model particles is given by \cite{fermion1}
\bea
\sigma v_{rel} &=& \frac{(g_S \sin \theta \cos \theta)^2}
                        {16\pi}
                   \left(1-\frac{4 m_{\psi}^2}{{s}}\right)
\nn\\
      & & \times \left(\frac{1}
            {({s}-m_{h_1}^2)^2 + m_{h_1}^2 \Gamma_{h_1}^2}
            +\frac{1}
               {({s}-m_{h_2}^2)^2 + m_{h_2}^2 \Gamma_{h_2}^2}\right.
\nn\\
      & & ~~~~~
      \left.+\frac{2({s}-m_{h_1}^2)({s}-m_{h_2}^2)
                    + 2 m_{h_1}m_{h_2}\Gamma_{h_1}\Gamma_{h_2}}
                  {(({s}-m_{h_1}^2)^2 + m_{h_1}^2 \Gamma_{h_1}^2)
                   (({s}-m_{h_2}^2)^2 + m_{h_2}^2 \Gamma_{h_2}^2)}
      \right)
\nn\\
      & & \times \left[
           \left(\frac{m_b}{v_0} \right)^2 \cdot 2 {s}
                 \left(1-\frac{4 m_b^2}{{s}}\right)^{3/2}\cdot 3
            + \left(\frac{m_t}{v_0} \right)^2 \cdot 2 {s}
              \left(1-\frac{4 m_t^2}{{s}}\right)^{3/2}\cdot 3
      \right.
\nn\\
      & & ~~~~~
           + \left(2\frac{m_W^2}{v_0}\right)^2
             \left(2+\frac{({s}-2m_W^2)^2}{4 m_W^4}\right)
                 \cdot\sqrt{1-\frac{4 m_W^2}{{s}}}
\nn\\
      & & ~~~~~
    \left. + \left(2\frac{m_Z^2}{v_0}\right)^2
             \left(2+\frac{({s}-2m_Z^2)^2}{4 m_Z^4}\right)
                 \cdot\sqrt{1-\frac{4 m_Z^2}{{s}}} \cdot \frac12
         \right]
\nn\\
      & &  + \sum_{i,j=1,2} \sigma_{h_i h_j}
           + \sum_{i,j,k=1,2} \sigma_{h_i h_j h_k} ,
\eea
where $\Gamma_{h_i}$ is the decay width of $h_i$ for $i=1,2$,
and $\sigma_{h_i h_j}$, $\sigma_{h_i h_j h_k}$ are
the annihilation cross sections of $\bar{\psi}\psi$ into $h_i h_j$
or $h_i h_j h_k$ with $i,j,k=1,2$. Now we can compute the thermal average of the cross section over $s$, $\langle \sigma v\rangle $, which is given by \cite{gondolo}
\be\label{thermal average}
\langle \sigma v\rangle=\frac{1}{8m^4T_FK^2_2(m/T_F)}\int_{4m^2}^\infty ds \sigma(s)(s-4m^2)\sqrt{s}K_1\left(\frac{\sqrt s}{T_F}\right),
\ee
where $K_{n}(x)$ is  the  modified Bessel function of order $n$. The freeze-out temperature $T_F$  is estimated by the following iterative equation:
 \be\label{tem}
 {x_F} = {\rm{ln}}\left( {\left. {\frac{m}{{2{\pi ^3}}}\sqrt {\frac{{45{M_{{\rm{pl}}}}}}{{2{g_*}{x_F}}}}
 \langle \sigma v\rangle } \right)} \right..
 \ee
where $x_F=m/T_F$ and ${M_{{\rm{pl}}}} = 1.22 \times {10^{19}}{\rm{ GeV}}$ is the Plank mass and ${g_*} = 91.5$ \cite{colb}.

\begin{figure}
\includegraphics[width=10cm]{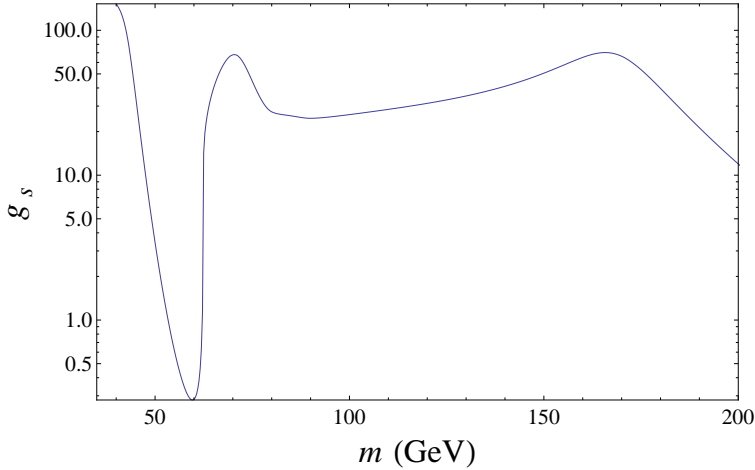}
\caption {\small
The figure shows the dependency of $g_s$ to $m$ with respect to the relic density condition.}
\label{fig2}
\end{figure}

We consider the SFDM with mass below 200 GeV which is the range of Fermi-Lat data in hand \cite{fermi}. Moreover, we take the SM Higgs mass 125 GeV which is consistent with the new reported value by ATLAS and CMS, and 500 GeV for the other Higgs.  Therefore, the annihilation of SFDMs into Higgs particles is not possible. There are a couple of free parameters in the thermally averaged annihilation cross section, in addition to the dark matter mass and Higgs masses: the coupling between the singlet fermion and singlet Higgs, $g_s$, the Higgses mixing parameter $\th$, and the coupling constants between singlet Higgs and the SM Higgs, $\lambda_1$ and $\lambda_2$.  The dependency of $\lambda_1$ and $\lambda_2$ occurs through the decay width of the Higgs messenger to the lighter Higgs in denominator, so in our desired range of energy these coupling constants are not relevant. This decay rate is also dependent on the mixing angle which is fixed in our calculation. Demanding the correct relic abundance for the dark matter we derive $g_s$ in terms of $m$.  Since in our calculations the cross sections directly depend on $g_s \sin(2\theta)$, we have fixed $\theta$ to conclude $g_s$ (One should note that the effect of the appearance $\theta$ in the Higgs decay width in the denominator is not very significant). Fig. \ref{fig2} illustrates $g_s$ vs dark matter mass with $\sin\theta=0.02$, for instance. It is clear that, if one considers a stronger mixing then the values of $g_s$ should be smaller. It is notable that,
  the valley represents the first resonance region where the dark matter mass is about half of the standard model Higgs mass. In addition, the decreasing region begins at about 170 GeV where the threshold of top quark production is.

\section{The $\bar\psi\psi\to\gamma\gamma$ cross section}\label{sec3}
\begin{figure}[h]
\vspace*{-5mm}
\[
  \vcenter{\hbox{
  \begin{picture}(130,80)(-20,0)
  \ArrowLine(5,15)(30,40)
  \Photon(75,55)(100,70){2}{5}
   \Photon(75,25)(100,10){2}{5}
  \DashArrowLine(30,40)(60,40){2}
  \Line(60,40)(75,55)
    \Line(60,40)(75,25)
    \Line(75,55)(75,25)
  \ArrowLine(30,40)(5,65)
   \Text(0,12)[b]{\scriptsize{$\psi$}}
  \Text(0,62)[b]{\scriptsize{$\bar\psi$}}
  \Text(45,43)[b]{\scriptsize{$H_j$}}
    \Text(70,38)[b]{\scriptsize{$t$}}
     \Text(105,8)[b]{\scriptsize{$\gamma$}}
  \Text(105,68)[b]{\scriptsize{$\gamma$}}
    \end{picture}
  \begin{picture}(230,80)(-40,0)
  \ArrowLine(5,15)(30,40)
  \Photon(75,55)(100,70){2}{5}
   \Photon(75,25)(100,10){2}{5}
  \DashArrowLine(30,40)(60,40){2}
  \DashLine(60,40)(75,55){4}
    \DashLine(60,40)(75,25){4}
    \DashLine(75,55)(75,25){4}
  \ArrowLine(30,40)(5,65)
  \Text(0,12)[b]{\scriptsize{$\psi$}}
  \Text(0,62)[b]{\scriptsize{$\bar\psi$}}
  \Text(45,43)[b]{\scriptsize{$H_j$}}
    \Text(70,38)[b]{\tiny{$W$}}
     \Text(105,8)[b]{\scriptsize{$\gamma$}}
  \Text(105,68)[b]{\scriptsize{$\gamma$}}
  \end{picture}}}
\]
\vspace*{-10mm}
\caption[]{The dominant Feynman diagrams for the annihilation of two singlet fermionic dark matter particles into monochromatic gamma ray lines.}
\label{fig1}
\end{figure}
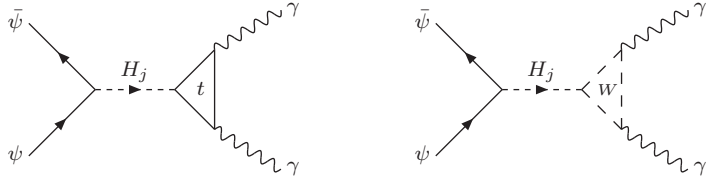

Now, according to the model presented in the previous section, we derive the cross section for the
 annihilation of two dark matter particles into gamma ray pairs. Although our dark matter cannot directly couple to the
  SM particles, the annihilation can occur by a Higgs particle through s-channel. In addition, since photon
  is a massless particle, it does not couple to the Higgs boson directly.
Nevertheless,  H$\gamma\gamma$ vertex, can be generated at the quantum level with loops involving massive particles,
such as W$^{\pm}$ bosons and
charged fermions.
The dominant Feynman diagrams contributing to this process are shown in Fig. \ref{fig1}.
\vskip 5pt
\begin{figure}[th]
\includegraphics[width=10cm]{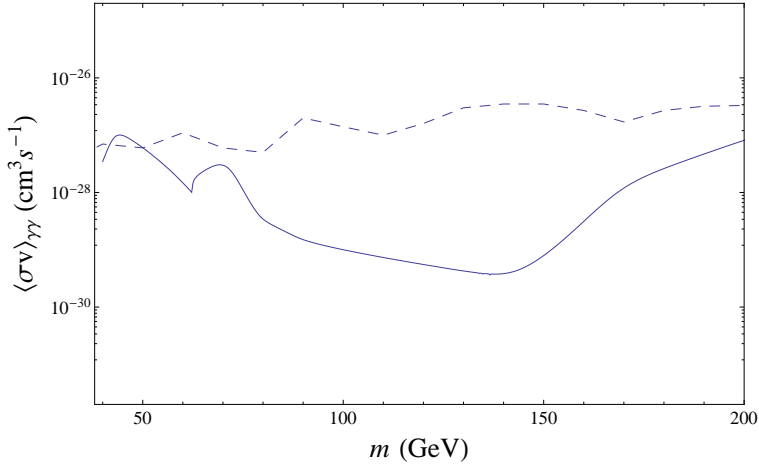}
\caption {\small
Annihilation cross section into gamma ray lines as a
function of dark matter mass for $m_{h_1}=125$  GeV, $m_{h_2}=500$  GeV,
 and $g_s$ coming from Fig. \ref{fig2}. Dashed line denotes the Fermi-Lat bound \cite{fermi}.
}
\label{fig4}
\end{figure}

Since the final products are massless, we can write down the cross section as follows:
\[\sigma v = \frac{1}{{8\pi s}}\frac{1}{4}\sum\limits_{\rm{spins}}
{{{\left| {{M_{\bar \psi \psi  \to \gamma \gamma }}} \right|}^2}} ,\]
where $\sqrt{s}$ is the energy in the center of mass frame, and
 \[{M_{\bar \psi \psi  \to \gamma \gamma }} =
  \sum\limits_{j = 1,2} {{{\bar v}}(p)(i{g_s}{S_j}\theta ){u}(p)} \frac{i}{{s - {m^2_{{h_j}}} - i{m_{{h_j}}}
  {\Gamma _j}}}{M_{{h_j} \to \gamma \gamma }},\]
 where ${S_1}\theta$ and ${S_2}\theta$
 denote $\sin \theta $ and $\cos \theta $, respectively,
 and ${M_{{h_j} \to \gamma \gamma }}$ is the amplitude for the decay of a Higgs into two photons. Here,
 ${\Gamma _j}$ is the decay width of $j$'th Higgs. One can write ${M_{{h_j}
  \to \gamma \gamma }}$ as follows \cite{shifman,gunion}:
\[{M_{{h_j} \to \gamma \gamma }} = \frac{{\sqrt {1 - ({S_i}{\theta})^2} \alpha gs}}
{{8\pi {M_W}}}[3{(\frac{2}{3})^2}{F_t} + {F_W}],\]
where
 \[{F_t} =  - 2\tau [1 + (1 - \tau )f(\tau )],\]
 \[{F_W} = 2 + 3\tau  + 3\tau (2 - \tau )f(\tau ).\]
Here, $\tau  = 4{m_i}^2/s$ with $i=t,W$ and
\[f(\tau ) = \left\{ \begin{array}{ll}
{\left( {{{\sin }^{ - 1}}\sqrt {1/\tau } } \right)^2},&\qquad{\rm{for}}\quad\tau  \ge {\rm{1   }}\\
 - \frac{1}{4}\left( {\left. {{\rm{ln}}\frac{{1 + \sqrt {1 - \tau } }}
 {{1 - \sqrt {1 - \tau } }} - i\pi } \right)^2} \right.&\qquad{\rm{for}}\quad\tau {\rm{ < 1 }.}
\end{array} \right.\]
Therefore, we obtain the cross section as follows:
\be
\begin{array}{l}
\sigma v=\frac{1}{8\pi s}\frac{1}{4}\sum\limits_{\rm{spins}} {{{\left| {{M_{\bar \psi \psi  \to \gamma \gamma }}}
 \right|}^2}}  = \frac{1}{32\pi s}{g_s}^2(s - 4{m^2})\left\{ {\frac{{{{\left| {{M_{{h_1}
 \to \gamma \gamma }}} \right|}^2}{{\sin }^2}\theta }}{{{{({m_{{h_1}}}^2 - s)}^2} +
 {m_{{h_1}}}^2{\Gamma _1}^2}}} \right. + \frac{{{{\left| {{M_{{h_2} \to \gamma \gamma }}}
  \right|}^2}{\rm{co}}{{\rm{s}}^2}\theta }}{{{{({m_{{h_2}}}^2 - s)}^2} + {m_{{h_2}}}^2{\Gamma _2}^2}}\\
\left. {{\rm{                     }} + \frac{{(s - {m_{{h_1}}}^2)(s - {m_{{h_2}}}^2)
 + \left| {{M_{{h_1} \to \gamma \gamma }}} \right|\left| {{M_{{h_2} \to \gamma \gamma }}} \right|}}
 {{\left[{{({m_{{h_1}}}^2 - s)}^2} + {m_{{h_1}}}^2{\Gamma _1}^2\right]\left[{{({m_{{h_2}}}^2 - s)}^2}
 + {m_{{h_2}}}^2{\Gamma _2}^2\right]}}\sin \theta {\rm{cos}}\theta ({M_{{h_1} \to \gamma \gamma }}
 {M^*_{{h_2} \to \gamma \gamma }} + {M^*_{{h_2} \to \gamma \gamma }}{M_{{h_1} \to \gamma \gamma }})} \right\}.
\end{array}
\ee
Thermal average of this cross section is also obtained using (\ref{thermal average}) and (\ref{tem}).
 We have illustrated the phenomenological aspects of this calculation via the Fig. \ref{fig4}.
  The SFDM can be served as cold dark matter provided that the free parameters are chosen such that the relic abundance condition is respected. Therefore, this figure shows cross section versus the dark matter mass while we use $g_s$ from Fig. \ref{fig2}.
    For comparison, we have also shown the upper limits obtained by Fermi-Lat \cite{fermi} for Navarro-Frenk-White (NFW) density profile.
It is noticeable that most of the region (except the resonance) is
actually excluded by direct detection experiments, following the results of
Figure 6 in Ref. \cite{fermion1}. However we should note that, for masses below about 70 GeV (and above about 150 GeV), in particular the resonance regions where are not excluded by direct detection experiment, the cross section is comparable with the Fermi-Lat data. Therefore, these values may  be explored by likely future more precise experiments.

\section{Summary and discussion}
In this paper we have considered an extension of the SM which includes a singlet fermion as a cold dark matter
 and a real singlet scalar as a messenger to explain the corresponding thermal relic density \cite{fermion1}. Demanding correct relic abundance, we have derived $g_s$  and plotted it in Fig. \ref{fig2} versus dark matter mass. This figure is consistent with figure 2 in ref. \cite{fermion1}. Only we have fixed the Higgs mixing angle to obtain a definite value for $g_s$.
Then the cross section of the annihilation of the singlet fermion into two monochromatic photons has been obtained. The thermal averaged of the annihilation of this cross section is calculated for a wide range of energies including resonance. We have depicted the thermally averaged of the SFDM annihilation cross section into two photons for a reasonable set of the Higgs masses in Fig. \ref{fig4}. This figure allows one to compare our result with the Fermi-Lat observations.
The most of the region (except the resonance) is excluded by direct detection experiments (see figure 6 in ref. \cite{fermion1}). However, the Fermi-Lat data can be an alternative way for exploring the nature of dark matter. As one can see from Fig. (\ref{fig4}), in some region (mass below 70 GeV and above 150 GeV which include the resonance regions), the two-photon annihilation cross section of SFDM is comparable to the resent Fermi-Lat data and potentially can be explored by likely future more precise experiments.

 {\bf Acknowledgement:} The financial
support of the University of Qom research council is acknowledged.

\end{document}